\documentclass[11pt]{article}
\usepackage{epsfig}
\usepackage{natbib}
\usepackage{amscd}
\usepackage{amsmath}
\usepackage{amsfonts}
\usepackage{amssymb}
\newcommand {\ctn}{\cite}

\newtheorem{theorem}{Theorem}

\newenvironment{proof}[1][Proof]{\textbf{#1.} }{\ \rule{0.5em}{0.5em}}

\begin{document}
\title{{\bf A Fully Bayesian Approach to Assessment of Model Adequacy in Inverse Problems}}
 \author{Sourabh Bhattacharya\footnote{Bayesian and Interdisciplinary Research Unit,
 Indian Statistical Institute, 203 B. T. Road, Kolkata-700108. Corresponding e-mail:~{\em sourabh@isical.ac.in.}}} 

\maketitle

\begin{abstract}
We consider the problem of assessing goodness of fit of a single Bayesian model
to the observed data in the inverse problem context. 
A novel procedure of goodness of fit test is proposed, based on construction of reference distributions 
using the `inverse' part of the given
model. This is motivated by an example from palaeoclimatology 
in which it is of interest to reconstruct past climates 
using information obtained from fossils deposited
in lake sediment. 
Since climate influences species, the model is built in the forward sense, that is, fossils are
assumed to depend upon climate. The model combines `modern data' which consists of observed species composition and the corresponding observed climates with
`fossil data'; the latter data consisting of fossil species composition deposited in lake sediments for the past thousands of years, but the corresponding past climates are unknown.  
Interest focuses on prediction of {\it unknown} past climates, which is the inverse part of the model.

Technically, given a model $f(Y\mid X,\theta)$, where $Y$ is the observed data and $X$ is a set of (non-random) covariates, we obtain
reference distributions based on the posterior $\pi(\tilde X\mid Y)$, where $\tilde X$ must be interpreted as the {\it unobserved} random vector corresponding to the
{\it observed} covariates $X$. 
Put simply, if the posterior distribution  $\pi(\tilde X\mid Y)$ gives high density to the observed covariates $X$, or equivalently, 
if the posterior distribution of $T(\tilde X)$ gives high density to $T(X)$, where $T$ is any appropriate statistic, then we say that the model fits the data. Otherwise the model
in question is not adequate. 
We provide decision-theoretic justification of our proposed approach 
and discuss other theoretical and computational advantages. 
We demonstrate our methodology with many simulated examples and three 
complex, high-dimensional, realistic palaeoclimate problems, including the motivating
palaeoclimate problem.

Although our proposal is ideally suited for checking model fit in inverse
regression problems, we indicate
that the proposal may be potentially extended for model checking in quite general Bayesian problems.
However, we do not claim to have solved all issues involved; 
in fact, our aim in this paper is to discuss advantages of, and also to shed light on issues that
could be potential future research topics. If nothing else, we hope to have been able to make a step forward
in the right direction.
\\[2mm]

\end{abstract}


\section{Introduction}
\label{sec:intro}

To quote \ctn{Gelman96}, assessing the plausibility of a posited model (or of assumptions in general) is always
fundamental, especially in Bayesian data analysis. 
Compared to the vast classical statistical literature that attempts to address the question of model assessment,
the Bayesian literature is much scarce. \ctn{Gelman96} seems to be the first to attempt extension of the essence of the classical approach
to the Bayesian framework. Their approach is based on computing the posterior distribution of the parameters given the data 
and then to compute a P-value, involving a discrepancy measure, which is a function of the data as well
as the parameters. Their approach differs from the available classical approaches mainly in introducing a discrepancy
measure that depends on the parameters as well. \ctn{Bayarri00} introduced two alternative P-values and demonstrated
that they are advantageous compared to the P-value of \ctn{Gelman96}.
In this paper, we introduce an approach based on `inverse reference distributions' (IRD). We argue that 
the approach is best suited for assessing Bayesian model fit in inverse problems but may be extended
to quite general Bayesian problems. The proposal is novel compared to 
the available approaches and has some distinct advantages.

The motivating example arises in quantitative palaeoclimate reconstruction where `modern data' consisting of multivariate counts of species
are available along with the observed climate values. Also available are fossil assemblages of the same species, but deposited in lake sediments 
for past thousands of years. This is the fossil species data. However, the past climates corresponding to the fossil species data are unknown, and it is of interest
to predict the past climates given the modern data and the fossil species data. Roughly, the species composition are regarded as functions
of climate variables, since in general ecological terms, variations in climate drives variations in species, but not vice versa.
However, since the interest lies in prediction of climate variables,  the inverse nature of the problem is clear. The past climates, which must be regarded as random variables,
may also be interpreted as {\it unobserved covariates}.  It is thus natural to put a prior probability distribution on the unobserved covariates.

Interestingly, the approach used for prediction of past climates motivates 
our Bayesian approach to assessment of model adequacy, in particular,
for inverse regression problems, using posterior distributions based on prior 
probability distributions on covariates, 
which are treated as unknown. Broadly, we say that the model fits the data if the posterior distribution of the random variables corresponding to the covariates
capture the observed values of the covariates. Otherwise, the model does not fit the data. It is worth noting that although the values of the covariates are known, we propose to fit 
the model {\it assuming that the values are unknown} and {\it predict the random variables that stand for the unknown covariates}. The covariates predicted in this manner can then be 
compared with the originally observed values to assess model fit in a fully Bayesian manner.  

The rest of our paper is structured as follows. In Section \ref{sec:overview} we review the existing literature 
on model assessment, all of which are
concerned with forward problems.
The key idea of our IRD approach is presented in Section \ref{sec:key}, and 
in Section \ref{sec:decision_theory} we provide a decision theoretic justification of our
proposed approach.
In Section \ref{sec:improper} we note that improper priors may render the reference posterior improper; 
in this context we suggest a remedy using cross-validation.
We provide a summary of our illustrations of the IRD approach with various examples
in Section \ref{sec:summary_simstudies}.
Application of our methodology to the motivating palaeoclimate
problem is discussed in Section \ref{sec:palaeo}. 
Conclusions and future work are discussed in Section \ref{sec:conclusions}.

Further details on methods, experiments and data analyses are provided in
the supplement \ctn{Bhatta12}, whose sections, figures and tables have the prefix
``S-'' when referred to in this paper. Here we briefly describe the contents of the supplement.
Some relevant discrepancy measures for IRD are provided 
in Section S-1 of the supplement. 
Further details regarding prior construction for the IRD approach in addition to that
presented in Section \ref{sec:improper}, are provided in Section S-2.
In Section S-3 
we provide a brief overview of 
Importance Resampling MCMC (IRMCMC) proposed by \ctn{Bhatta06a},
for cross-validation in inverse problems, which is an indispensable computational method for IRD.
The complete details of the summary of the illustrations of IRD with simulation examples
outlined in Section \ref{sec:summary_simstudies}, are presented
in Section S-4. In Sections S-5 and S-6 we discuss applications of IRD to extensions of the
motivating palaeoclimate problem presented in Section \ref{sec:palaeo}.

\section{Overview of methods of model assessment in forward problems and their limitations}
\label{sec:overview}

One approach for checking the fit of a model is by examining the marginal distribution of 
the data (\ctn{Box80}).
Specifically, if
the marginal density of $Y$
is small, then
$Y$ is unlikely under the given model.
A problem with this approach is that, for improper prior on the parameter $\theta$, the marginal 
is improper.
Another problem is to decide on precisely how small the marginal
should be so that $Y$ can be treated as unlikely to
lead to rejection of the model.
The approach of reference distributions may be applied to this idea, but the problem
of impropriety of the marginal for improper priors is an impediment. Cross-validation
may be used as a proxy for the marginal, but in this case, strictly speaking, data $Y$
would be used twice; once to compute the cross-validation posteriors $\{\pi(\cdot\mid X,Y_{-i});i=1,\ldots,n\}$
and again to construct the discrepancy measure.

\ctn{Gelman96} recommended 
generalised test statistics $T(Y,\theta)$ that depend
on the parameters as well as the data, and proposed a Bayesian P-value for assessing goodness of fit. 
However, their model checking
strategy uses the data twice: once to compute the observed statistic $T(Y,\theta)$ and again to
obtain the posterior predictive reference distribution. 
\ctn{Bayarri00} demonstrated with examples that using data twice is undesirable; see also \ctn{JKG05}.
Specifically, in such cases, even with arbitrarily strong evidence against the null model,
the P-value does not tend to zero.
Also, the posterior predictive P-values do not generally have a uniform distribution under
the null hypothesis, not even
asymptotically (\ctn{Bayarri00}, \ctn{Robins00}). 

\ctn{Bayarri00} developed a related
approach based on posterior distributions that condition on only part of the information
in the data rather than using the full posterior distribution  
to define the reference distribution.
\ctn{Robins00} showed that their P-values are asymptotically uniformly distributed under the null hypothesis. 
But it is not clear to the author of this paper (see also the discussion by Evans of the paper by Bayarri and Berger) if the same is true 
for a finite sample size.
Another important point is that, given a specific discrepancy measure computation
of their P-value is burdensome and requires knowledge of the analytic form of the density of the specific discrepancy measure, which is not available
in general. Arguments are provided in \ctn{Bayarri00} that estimation of the density of a particular discrepancy measure is not difficult; however,  since
the above authors did not provide
any guidelines how to choose the right discrepancy measure, many possible discrepancy measures
must be considered. But then computation of P-values for each discrepancy measure has to
be done afresh; this will certainly become computationally very expensive for complex problems. \ctn{Stern00} pointed out that
the approach of \ctn{Bayarri00} can be very difficult to apply for the kinds of complex models that are most challenging to check in practice.
\ctn{Bayarri00} demonstrated their proposal with many theoretical examples,    
but they did not provide assessment of the performance of their methodology in the case of complex, real problems.

\section{The key idea of IRD}
\label{sec:key}
The essential idea of constructing an IRD can be described as follows. Suppose that data $Y=\{y_i\},i=1,\ldots,n$ are available.
Also available, suppose, are covariates $X=\{x_i\},i=1,\ldots,n$. Each of $y_i$ or $x_i$ may also be multivariate. We assume that there is a probability model associated with $Y$, given covariates $X$. We also assume,
as is natural, that $X$ is not associated with any probability model. So, we treat $Y$ as the data, but $X$ as fixed constants.  
To proceed with our approach we first pretend that the values of the covariates are unknown, probabilistically interpreted as random variables, which we denote as $\tilde X$. This unknown set of random
variables $\tilde X$, which may also be thought of as a replicate of the observed covariates $X$, must be predicted from data $Y$, in  an inverse sense. If the predicted values of $\tilde X$ are consistent with observed $X$
then we say that the model fits the data adequately, otherwise we say that the model does not fit the data.
A fully Bayesian approach to this prediction problem requires computation of an inverse reference distribution
based on the posterior 
\begin{equation}
\pi(\tilde X\mid Y)\propto\int \pi(\tilde X,\theta)L(Y,\tilde X,\theta)d\theta
\label{eq:inv_reference}
\end{equation}
In the above, $L$ denotes the likelihood of the unknowns $(\tilde X,\theta)$ where $\theta$ is the set of model parameters. It is important to observe that the above posterior does 
not depend upon the observed covariates $X$;  in other words, the model is
fitted without using the observed covariates. In the above posterior both $\theta$ and $\tilde X$ can be regarded as unknown
parameters. In our approach the model parameter $\theta$ will be regarded as a set of nuisance parameters 
(more discussion to follow subsequently) and $\tilde X$
will be regarded as the parameters of interest. The prior on $\tilde X$ and $\theta$
has been denoted by $\pi(\tilde X,\theta)$. We discuss in this paper that, 
based on whether or not observed $X$ is supported by the above posterior, an effective overall goodness-of-fit test, which
has some desirable properties, can be devised. 

Observe that in our proposal, the data is not used twice, since computation of the posterior (\ref{eq:inv_reference}) involves
conditioning on $Y$ alone.  
The observed covariates will be used only for the construction of the discrepancy measure. 
Denoting by $T(X)$ a discrepancy measure involving only observed covariates $X$, we construct the corresponding
reference distribution of the random discrepancy measure $T(\tilde X)$. Some examples of discrepancy measures are 
provided in Section S-1 of the supplement; for applications in this paper we throughout use the discrepancy measure (1) of Section S-1,
given by
\begin{equation}
T(X) = T_1(X) = \sum_{i=1}^n\frac{(x_i - E_{\pi}(\tilde x_i))^2}{V_{\pi}(\tilde x_i)},
\label{eq:discrepancy_T1}
\end{equation}
where $E_{\pi}$ and $V_{\pi}$ denote the mean and the variance with respect to $\pi(\cdot\mid Y)$.
Then if $T(X)$ lies within the appropriate credible region of $T(\tilde X)$, the model will be accepted, otherwise it will be
rejected. This we formalize decision theoretically in Section \ref{sec:decision_theory}. 
Thus, unlike other approaches (both Bayesian and classical), 
we have clearly defined a method that can decide 
whether to accept or to reject the model in question. An important issue to address in this context is whether or not the discrepancy measure 
should be allowed to depend upon the model parameters $\theta$. In the forward context, \ctn{Gelman96} defined general discrepancy measures
that depend upon both data and the model parameters. However, in our inverse approach letting the discrepancy measure depend upon the model
parameters will often not be meaningful, unless we let the discrepancy measure also depend upon $Y$. But this would imply using data $Y$ twice; once
to compute the posterior (\ref{eq:inv_reference}) and again to compute the discrepancy measure. So we strongly recommend that discrepancy measures
be independent of the model parameters.  We discuss this with an example.  Let us consider a  Poisson regression model, which we will use to illustrate our proposal, 
given by $y_i\sim Poisson(\theta x_i);i=1,\ldots,n$.
In the forward context, a discrepancy measure based on the residuals $y_i-\theta x_i$ seems natural. However, in our inverse approach an analogous measure
is not permissible, since this would entail using $Y$ twice, as indicated above. 
Indeed, one of our aims is to avoid using double use of the data. Moreover, in the case of complex hierarchical models there may be thousands of model parameters
and in such cases it is not clear how to construct a sensible discrepancy measure using such high-dimensional model parameter. In our opinion, it makes more sense
to integrate out the model parameters and base the discrepancy measure solely on the covariates. In other words, we treat the model parameters as nuisance parameters
in our approach. For details regarding nuisance parameters see \ctn{Berger99}.   


In the next section we formalize our proposed approach by providing
a decision theoretic justification. Based on ``0-1" loss function, we also provide a simple, but explicit, formula
for accepting or rejecting the model in question.

\section{Decision theoretic justification of our proposed IRD approach}
\label{sec:decision_theory}

If the data really come from the model assumed, then for any general discrepancy measure $T$, $T(\tilde X)$
is expected to give high probability density to the point $T(X)$.
In other words, we say that the model does not fit the data if for some 
pre-assigned quantity $\epsilon$,
\[\Bigg|\frac{T(\tilde X)-T(X)}{\sqrt{V_{\pi}(T(\tilde X)\mid Y)}}\Bigg| >\epsilon\]
with high posterior probability. 
Here we remind the reader that $\tilde X$ is to be considered a set of random variables or unknown parameters
corresponding to the true values $X$. The random and observed discrepancy measures $T(\tilde X)$
and $T(X)$ can likewise be treated as a parameter and the true value of the parameter respectively.
Using this framework, it is easy to formulate a Bayesian hypothesis testing problem, in the spirit
of that provided in \ctn{Berger85}. Note that we could not do the same if the discrepancy measure were dependent on data $Y$; this is
because $Y$ is the data arisen from a probability model and can not be interpreted as parameter. Since all other available approaches
to model assessment use discrepancy measures involving data $Y$, they do not have the Bayesian decision theoretic framework.

To put it formally, we are interested in testing
\[H_0 :\hspace{3mm}\Bigg|\frac{T(\tilde X)-T(X)}
{\sqrt{V_{\pi}(T(\tilde X)\mid Y)}}\Bigg|\leq\epsilon\]
against
\[H_1 :\hspace{3mm}\Bigg|\frac{T(\tilde X)-T(X)}
{\sqrt{V_{\pi}(T(\tilde X)\mid Y)}}\Bigg|>\epsilon\]
For $k=0,1$, let $\mathcal T_k$ denote the parameter space of $T(\tilde X)$ implied
by $H_k$. 
We denote acceptance of $H_k$ by $a_k$ 
and consider the ``0-1'' loss function
$\mathcal L(T(\tilde X),a_{k}) = 0$ if $T(\tilde X)\in\mathcal T_{k}$
and 
$\mathcal L(T(\tilde X),a_{k}) = 1$ if $T(\tilde X)\in\mathcal T_{\ell}; \ell\neq k$.

Then Bayes action (see \ctn{Berger85} for definition) is simply that for which the posterior expected loss 
$E_{\pi(\cdot\mid Y)}\{\mathcal L(T(\tilde X),a_{k})\}$; $k = 0, 1$
is the smaller, which implies that the Bayes decision is simply the hypothesis
with larger posterior probability. 
If 
\begin{equation}
p = \pi\left(\Bigg|\frac{T(\tilde X)-T(X)}
{\sqrt{V_{\pi}(T(\tilde X)\mid Y)}}\Bigg|\leq\epsilon\mid Y\right), 
\label{eq:model_probability}
\end{equation}
then $H_0$ is to be accepted if $p>1-p$, i.e. $p>1/2$.  It is important to note that the posterior probability is not a P-value, 
nor is it related to any P-value of any kind. It is simply a posterior probability of a unknown parameter. Also, very clearly, the data is not used
twice in computing the posterior probability. Due to this reason the probability given by (\ref{eq:model_probability})
has appropriate behaviour under the null hypothesis; in fact, very clearly, the posterior probability has 
uniform distribution for any size of the data.
It is useful to briefly clarify in this context the issue
of double-use of the data and the consequences. The posterior predictive P-value of \ctn{Gelman96} is defined as
\[P_{post}(Y)=\int Pr(T(\tilde Y,\theta)>T(Y,\theta)\mid\theta)\pi(\theta\mid Y)d\theta.\]
The (frequentist) distribution of $P_{post}(Y)$ depends upon the distribution of the conditioned ``data" $Y$.
Since, $Y$ is used twice, the P-value is ``overconfident" and the distribution is not even 
asymptotically $Uniform(0,1)$.
Now consider the posterior probability that corresponds to the IRD approach, as below:
\[P_{IRD}(X,Y)=Pr(T(\tilde X)>T(X)\mid Y).\]
The theorem below asserts that $P_{IRD}(X,Y)$ follows $Uniform(0,1)$ for any sample size under the distributions of $X$ and the marginal distribution of $Y$.

\begin{theorem}
\label{theorem:IRD_unif1}
Let $\tilde X\sim\pi(\tilde X)$ and $\theta\sim\pi(\theta)$, and let the priors be proper. The marginal distribution of $Y$ is given by 
$f(Y)=\int[Y|\tilde X,\theta]\pi(\tilde X)\pi(\theta)d\tilde Xd\theta$.
Then, for any sample size $n\geq 1$, $P_{IRD}(X,Y)\sim Uniform(0,1)$, with respect to $X\stackrel{\mathcal L}{\equiv}\tilde X$ and $Y\sim f(Y)$.
\end{theorem}
\begin{proof}
Let $F(T(X)|Y)$ denote the distribution function of $\pi\left(T(\tilde X)|Y\right)$, evaluated at $T(X)$. 
If $X\stackrel{\mathcal L}{\equiv}\tilde X$, then $F(T(X)|Y)\sim Uniform (0,1)$, almost surely with respect to $Y\sim f(Y)$.
In other words, if $X\stackrel{\mathcal L}{\equiv}\tilde X$, for almost all $Y\sim f(Y)$,
\begin{equation}
\int_{F(T(X)|Y)\geq 1-\gamma}\pi(X|Y)dX=\gamma.
\label{eq:IRD_unif1}
\end{equation}
Then note that if $X\stackrel{\mathcal L}{\equiv}\tilde X$, for any $\gamma\in (0,1)$, 
\begin{align}
	P_{X,Y}\left(P_{IRD}(X,Y)\leq\gamma\right)&=P_{X,Y}\left(F(T(X)|Y)\geq 1-\gamma\right)\notag\\
	&=\int\left[\int_{F(T(X)|Y)\geq 1-\gamma}\pi(X|Y)dX\right]f(Y)dY\notag\\
	&=\gamma\int f(Y)dY~~(\mbox{due to}~(\ref{eq:IRD_unif1}))\notag\\
	&=\gamma,\notag
\end{align}
proving the theorem.
\end{proof}

\subsection{Simulation experiment with respect to Theorem \ref{theorem:IRD_unif1}}
\label{subsec:simexp1}
Consider $y_i\sim Poisson(\theta x_i)$, $i=1,\ldots,n$, where $Poisson(\theta x_i)$ indicates Poisson distribution with mean $\theta x_i$.
Let $\pi(\theta)\equiv Uniform(0,10^5)$, and $\pi(\tilde X)=\prod_{i=1}^n\exp(-\tilde x_i)$. For $n=5$, we numerically compute the distribution of $P_{IRD}(X,Y)$,
after simulating $(X,Y)$ $1000$ times and obtaining $P_{IRD}(X,Y)$ using $1000$ draws from $\pi(\tilde X|Y)$. The results summarized in Table \ref{table:simexp1}
demonstrates that indeed $P_{IRD}(X,Y)\sim Uniform(0,1)$.
\begin{table}
\caption{Simulation experiment demonstrating that $P_{IRD}(X,Y)\sim Uniform(0,1)$.}
\label{table:simexp1}
\begin{center}
\begin{tabular}{|c||c|}\hline
$\gamma$ & $Pr\left(P_{IRD}(X,Y)\leq\gamma\right)$\\
\hline
0.0 & 0.000\\
0.1 & 0.084 \\
0.2 & 0.193 \\
0.3 & 0.297 \\
0.4 & 0.390 \\
0.5 & 0.487 \\
0.6 & 0.605 \\
0.7 & 0.713\\
0.8 & 0.800\\
0.9 & 0.897\\
1.0 & 1.000\\
\hline
\end{tabular}
\end{center}
\end{table}

Thus, the decision theoretic
framework formally shows that our proposed approach is fully Bayesian with a solid theoretical justification. Other available
model checking methods, which are all based on the forward part do not have appropriate calibration properties, at least if the size of the data set is finite. 
 \ctn{Sellke01}
attempt to provide calibration of P-values, but that is a ``lower bound" calibration  which may be too low,
especially for larger sample sizes (see the rejoinder of \ctn{Bayarri00}).
In the case of forward problems, \ctn{Hjort06} proposed a method of calibration,  but the method 
seems to work only if the prior distribution of the model parameters is proper. Hence, although the proposal of \ctn{Hjort06}
is promising, given that improper prior distribution is very widely used,
it is also useful to seek alternative criteria.

\subsection{Choice of $\epsilon$}
\label{subsec:choice_epsilon}
The choice of $\epsilon$ may be subjective, differing from problem to problem.
However, under the true model, we would expect the predicted values
of $\tilde X$ (which may be posterior means, medians, modes, etc.) to be close to observed $X$. 
Thus, under the true model, $T(X) \approx 0$ and we would expect 
\begin{equation}
\Bigg|\frac{T(\tilde X)-T(X)}{\sqrt{V_{\pi}(T(\tilde X)\mid Y)}}\Bigg| 
\approx\frac{|T(\tilde X)|}{\sqrt{V_{\pi}(T(\tilde X)\mid Y)}}
\label{eq:percentile}
\end{equation}
under $H_0$. 
Hence percentiles of the random variable on the right hand side of (\ref{eq:percentile})
may be reasonable choices of $\epsilon$. In other words, as a rule of thumb, for a particular choice
of $\alpha\in (0,1)$, $\epsilon$ may be regarded as the $(1-\alpha)$-th percentile of 
$\frac{|T(\tilde X)|}{\sqrt{V_{\pi}(T(\tilde X)\mid Y)}}$.
In this paper we illustrate the power of our test by observing if the observed discrepancy measure 
$T(X)$ falls within
the relevant $100(1-\alpha)=97\%$ credible region of the above reference distribution. 
We have, however, experimented with
several other sizes of the credible region corresponding to $T(\tilde X)$, but our main conclusions 
remained unchanged.

\section{Impropriety of the reference distribution and remedy using cross-validation}
\label{sec:improper}
Note that the integrand of (\ref{eq:inv_reference}) involves
the model parameters $\theta$ as well as $\tilde X$. Thus there will be more random variables than the number of data points
if each of $x_i$ and $y_i$ are of same dimensionality. In this case, improper prior on any unknown, 
$\tilde X$ or $\theta$ (the prior on $\tilde X$, if empirically estimated from $X$, may be proper, but the
prior on $\theta$ will often be taken as improper in complex hierarchical Bayesian problems), will  make the
posterior distribution $\pi(\tilde X,\theta\mid Y)$ improper if the data fails to provide information on 
that unknown. 
Using the already introduced Poisson regression model, we illustrate the issue of impropriety
of the joint posterior if $\theta$ has improper prior. 

Assume $y_i\sim Poisson(\theta x_i); i=1,\ldots,n$ where there are $n$ data points but $n+1$ unknowns 
in $\tilde X$ and $\theta$. Let us consider a proper prior for $\tilde X$, given by
\begin{equation}
\pi(\tilde X)\propto\prod_{i=1}^n\exp(-\beta\tilde x_i){\tilde x_i}^{\alpha-1}
\label{eq:proper_prior_x}
\end{equation}
As for $\theta$, suppose we use a uniform improper prior $\pi(\theta)=1$ for all $\theta$.
Then, the joint posterior of $\tilde X$ and $\theta$ is 
\begin{equation}
\pi(\tilde X,\theta\mid Y)\propto \exp\left\{-(\theta+\beta)\sum_{i=1}^n\tilde x_i\right\}
\theta^{\sum_{i=1}^ny_i}
\prod_{i=1}^n{\tilde x_i}^{y_i+\alpha-1}
\label{eq:posterior_x_theta}
\end{equation}
The marginal posterior of $\theta$ is given by
\begin{equation}
	\pi(\theta\mid Y)\propto\frac{\theta^{\sum_{i=1}^n y_i}}{(\theta+\beta)^{(\sum_{i=1}^ny_i+n\alpha)}}
\label{eq:marg_post_theta}
\end{equation}
However,
\begin{equation}
\int_0^{\infty} \frac{\theta^{\sum_{i=1}^n y_i}}{(\theta+\beta)^{(y_i+\alpha)}}d\theta
=\int_0^1 z^{\alpha n-2}(1-z)^{\sum_{i=1}^n y_i}dz
\label{eq:beta_integration}
\end{equation}
and the above integration converges if and only if $n>\frac{1}{\alpha}$. In other words,
if the prior on $\tilde X$ is vague, which is signified by $\alpha\approx 0,\beta\approx 0$, then 
the data size $n$ has to be impractically large to render the posterior of $\theta$ proper.
Hence, in general, for this problem, 
\begin{equation}
\int \pi(\theta\mid Y)d\theta=\infty 
\label{eq:improper_marg_theta}
\end{equation}
Now observe that,
\begin{equation}
\int\int\pi(\tilde X,\theta\mid Y)d\tilde X d\theta
=\int\left\{\int\pi(\tilde X\mid Y,\theta)d\tilde X\right\}\pi(\theta\mid Y)d\theta
\label{eq:improper_joint}
\end{equation}
Note, that in (\ref{eq:improper_joint}), for each $\theta$, $\pi(\tilde X\mid Y,\theta)$ is proper;
in fact, just a product of proper Gamma densities
(can easily be seen from (\ref{eq:posterior_x_theta})), so $\int\pi(\tilde X\mid Y,\theta)d\tilde X=1$.
However, since the term $\int\pi(\theta\mid Y)d\theta=\infty$ by (\ref{eq:improper_marg_theta}),
the above integral (\ref{eq:improper_joint}) is infinity as well. Hence, the joint posterior
$\pi(\tilde X,\theta\mid Y)$ is improper.

We note, however, that it is not necessarily the case that the distribution of $T(\tilde X)$ will
be improper if the posterior $\pi(\tilde X\mid Y)$
is improper (to consider a pedagogical example, if $\pi(\psi)=1;\psi\in (0,\infty)$ then $\exp(-\psi)$ has a 
proper distribution on $(0,1)$). 
However, in general, the distribution of $T(\tilde X)$ will be analytically intractable, 
and it must be obtained using MCMC simulations
of $(\tilde X,\theta)$ from the joint posterior $\pi(\tilde X,\theta\mid Y)$ (since the marginal posterior 
distribution $\pi(\tilde X\mid Y)$ is analytically intractable). Now, if the joint posterior
$\pi(\tilde X,\theta\mid Y)$ is improper, then MCMC simulations from this posterior
will not make sense. 
Hence, it is very important that the joint posterior is proper.

To avoid the problem of impropriety, we propose to approximate
the true posterior distribution using cross-validation. In other words, we propose to simulate
from the leave-one-out posteriors $\{\pi(\tilde x_i\mid X_{-i},Y);i=1,\ldots,n\}$, where $X_{-i}$ stands for the data, omitting
in each case the corresponding $x_i$. The random variable $\tilde x_i$ corresponds to the omitted value $x_i$. Note that
the leave-one-out posterior with case $i$ omitted is given by
\begin{equation}
\pi(\tilde x_i\mid X_{-i},Y)\propto\int \pi(\tilde x_i,\theta)f(y_i\mid \tilde x_i,\theta)\prod_{j\neq i}f(y_j\mid x_j,\theta)d\theta
\label{eq:xval}
\end{equation}
In the above integrand, $\tilde x_i$ and $\theta$ are the only random variables, whereas $Y$ is the data and $x_j;j\neq i$
are known constants. Hence there are much less unknowns compared to the number of knowns; this usually results
in a proper posterior of $\tilde x_i$ and $\theta$. With the above Poisson regression example, but with
improper priors on both $\tilde X$ and $\theta$, that is, $\pi(\tilde x_i)=1$; $x_i>0$ for $i=1,\ldots,n$
and $\pi(\theta)=1$; $\theta>0$,
it can be shown that the cross-validation posterior of $\theta$, with $x_i$ deleted, is given by
\begin{equation}
\pi(\theta\mid X_{-i},Y)=\frac{\left(\sum_{j\neq i}x_j\right)^{\left(\sum_{j\neq i}y_j\right)}}
{\Gamma(\sum_{j\neq i}y_j)}
\theta^{\left(\sum_{j\neq i}y_j-1\right)}\exp\left\{-\theta\sum_{j\neq i}x_j\right\},
\label{eq:cv_post_theta}
\end{equation}
which is a Gamma distribution.
The cross-validation posterior of $\tilde x_i$, when $x_i$ is deleted, is given by
\begin{equation}
\pi(\tilde x_i\mid X_{-i},Y)=\frac{\left(\sum_{j\neq i}x_j\right)^{\left(\sum_{j\neq i}y_j\right)}
\Gamma\left(\sum_{j=1}^ny_j+1\right)}{\Gamma(y_i+1)\Gamma(\sum_{j\neq i}y_j)}
\frac{{\tilde x}^{y_i}}
{\left(\tilde x+\sum_{j\neq i}x_j\right)^{\left(\sum_{j=1}^ny_j+1\right)}}
\label{eq:cv_post_x}
\end{equation}
Clearly, both the marginal cross-validation posteriors (\ref{eq:cv_post_theta})
and (\ref{eq:cv_post_x}) are proper, although
the priors of $\theta$ and $\tilde x_i$ are improper. Certainly, 
unlike in the case of (\ref{eq:improper_joint}), 
the joint posterior 
$\pi(\tilde x_i,\theta\mid X_{-i},Y)$ is proper. Hence, MCMC simulation from $\pi(\tilde x_i,\theta\mid X_{-i},Y)$
is not problematic, even though the priors of both $\tilde x_i$ and $\theta$ are improper.

Observe that the set of leave-one-out posteriors $\{\pi(x_i\mid X_{-i},Y);i=1,\ldots,n\}$ is equivalent
to $\pi(x_1,\ldots,x_n\mid Y)$. To see this, let $X_0=(x_{10},\ldots,x_{n0})'$ be any fixed point in the support of 
$\pi(x_1,\ldots,x_n\mid Y)$. Then it holds that
\begin{eqnarray}
\pi(x_1,\ldots,x_n\mid Y)&=&\frac{\pi(x_1\mid x_2,\ldots,x_n,Y)}{\pi(x_{10}\mid x_2,\ldots,x_n,Y)} \frac{\pi(x_2\mid x_{10},x_3\ldots,x_n,Y)}{\pi(x_{20}\mid x_{10},x_3,\ldots,x_n,Y)}\nonumber\\
                                           &\ldots& \frac{\pi(x_n\mid x_{10},\ldots,x_{n-1,0},Y)}{\pi(x_{n0}\mid x_{10},\ldots,x_{n-1,0},Y)}\pi(x_{10},\ldots,x_{n0}\mid Y)
                                           \label{eq:brooks_lemma}
\end{eqnarray}                                           
 This follows from Brook's lemma (\ctn{Brook64}), the usefulness of which is exposed in \ctn{Besag74}. Equation (\ref{eq:brooks_lemma}) expresses
 the joint distribution 
 $\pi(x_1,\ldots,x_n\mid Y)$ in terms of the full conditional distributions. Note that  the factor  $\pi(x_{10},\ldots,x_{n0}\mid Y)$  appearing on the right-hand side of (\ref{eq:brooks_lemma}) 
 is just a constant. Hence the joint distribution is determined by the full conditional distributions up to a proportionality constant.

So, under the true model
a draw from each of $\pi(\cdot\mid X_{-i},Y);i=1,\ldots,n$, will {\it approximate} $\pi(x_1,\ldots,x_n\mid Y)$.
In particular, even if the true posterior distribution $\pi(x_1,\ldots,x_n\mid Y)$ is improper, the 
approximating posterior
distribution induced by the cross-validation posteriors, will be proper (see also \ctn{Gelfand96}, \ctn{Carlin00}).
In fact, in such case, $\{x_1,\ldots,x_n\}$ can be looked upon as just a realisation from $\pi(\cdot\mid Y)$.
One might argue that strictly speaking, $X$ is being used twice; once to compute the posteriors and again
to construct the discrepancy statistic. However, we had argued earlier that $X$
may not be treated as the data since there is no probability model associated with it. Even so, if $(X,Y)$ is considered the entire data set,
then since $Y$ is not used to compute the discrepancy measure, the entire data set is not used twice in our implementation. 
In problems where there are no impropriety issues, our experiments revealed (not reported in this paper) that the results obtained by directly
computing (\ref{eq:inv_reference}) are equivalent to the results obtained by implementing this cross-validation idea. 
This is also confirmed by the following theorem, which asserts that the relevant probability $P_{IRD}(X,Y)$ computed using cross-validation,
follows the $Uniform(0,1)$ distribution.
\begin{theorem}
\label{theorem:IRD_unif2}
Let $\tilde X\sim\pi(\tilde X)$ and $\theta\sim\pi(\theta)$, and let the priors be proper. The marginal distribution of $Y$ is given by 
$f(Y)=\int[Y|\tilde X,\theta]\pi(\tilde X)\pi(\theta)d\tilde Xd\theta$. Assume that $P_{IRD}(X,Y)$ is computed based on the cross-validation posteriors
$\left\{\pi(\tilde x_i|X_{-i},Y):i=1,\ldots,n\right\}$.
Then, for any sample size $n\geq 1$, $P_{IRD}(X,Y)\sim Uniform(0,1)$, with respect to $X\stackrel{\mathcal L}{\equiv}\tilde X$ and $Y\sim f(Y)$.
\end{theorem}
\begin{proof}
Let $\tilde\pi(\tilde x_1,\ldots,\tilde x_n|X,Y)=\prod_{i=1}^n\pi(\tilde x_i|X_{-i},Y)$.	
Let $\tilde F(T(X)|X,Y)$ denote the distribution function of $\tilde\pi\left(T(\tilde X)|X,Y\right)$, evaluated at $T(X)$. 
Note that the distribution function $\tilde F(T(\tilde X)|X,Y)$ is computed based on the cross-validation posteriors $\left\{\pi(\tilde x_i|X_{-i},Y):i=1,\ldots,n\right\}$.
Hence, it is clear that the distribution of $\tilde F(T(X)|X,Y)$ is induced by $\left\{\pi(x_i|X_{-i},Y):i=1,\ldots,n\right\}$.
Now recall from Brook's lemma that $\left\{\pi(x_i|X_{-i},Y):i=1,\ldots,n\right\}$ is equivalent to $\pi(X|Y)$.
It follows that, if $X\stackrel{\mathcal L}{\equiv}\tilde X$, then $F(T(X)|X,Y)\sim Uniform (0,1)$, almost surely with respect to $Y\sim f(Y)$.
In other words, if $X\stackrel{\mathcal L}{\equiv}\tilde X$, then for almost all $Y\sim f(Y)$,
\begin{equation}
\int_{\tilde F(T(X)|X,Y)\geq 1-\gamma}\pi(X|Y)dX=\gamma.
\label{eq:IRD_unif2}
\end{equation}
Then note that if $X\stackrel{\mathcal L}{\equiv}\tilde X$, then for any $\gamma\in (0,1)$, 
\begin{align}
	P_{X,Y}\left(P_{IRD}(X,Y)\leq\gamma\right)&=P_{X,Y}\left(\tilde F(T(X)|X,Y)\geq 1-\gamma\right)\notag\\
	&=\int\left[\int_{\tilde F(T(X)|Y)\geq 1-\gamma}\pi(X|Y)dX\right]f(Y)dY\notag\\
	&=\gamma\int f(Y)dY~~(\mbox{due to}~(\ref{eq:IRD_unif2}))\notag\\
	&=\gamma,\notag
\end{align}
proving the theorem.
\end{proof}

\subsection{Simulation experiment with respect to Theorem \ref{theorem:IRD_unif2}}
\label{subsec:simexp2}
As in Section \ref{subsec:simexp1} consider $y_i\sim Poisson(\theta x_i)$, $i=1,\ldots,n$, with $\pi(\theta)\equiv Uniform(0,10^5)$, 
and $\pi(\tilde X)=\prod_{i=1}^n\exp(-\tilde x_i)$. For $n=5$, we numerically compute the distribution of $P_{IRD}(X,Y)$ using cross-validation,
after simulating $(X,Y)$ $1000$ times and obtaining $P_{IRD}(X,Y)$ using $1000$ draws from $\left\{\pi(\tilde x_i|X_{-i},Y):i=1,\ldots,n\right\}$. 
The results summarized in Table \ref{table:simexp2}
demonstrates that cross-validation based $P_{IRD}(X,Y)\sim Uniform(0,1)$.
\begin{table}
\caption{Simulation experiment demonstrating that cross-validation based $P_{IRD}(X,Y)\sim Uniform(0,1)$.}
\label{table:simexp2}
\begin{center}
\begin{tabular}{|c||c|}\hline
$\gamma$ & $Pr\left(P_{IRD}(X,Y)\leq\gamma\right)$\\
\hline
0.0 & 0.001\\
0.1 & 0.104 \\
0.2 & 0.203 \\
0.3 & 0.319 \\
0.4 & 0.419 \\
0.5 & 0.509 \\
0.6 & 0.605 \\
0.7 & 0.705\\
0.8 & 0.798\\
0.9 & 0.894\\
1.0 & 1.000\\
\hline
\end{tabular}
\end{center}
\end{table}

\subsection{Further discussion regarding our cross-validation based IRD approach}
\label{subsec:further_dicsussion_xval}
Since in a very large class of Bayesian
models the impropriety problem will arise, for the sake of generality we recommend this cross-validation 
idea for implementation of model adequacy test.
Moreover, cross-validation has a nice intuitive appeal, and can provide insight into finer aspects of 
the data in addition to providing
an overall goodness of fit statistic. For instance, it can be checked if any individual $x_i$ is an outlier with respect to the Bayesian model; for details, see Section S-1 of the supplement. Also,
by noting the number of observed $x_i$ falling within the respective credible regions it is possible to obtain more information about model fit issues. This is exactly the procedure we use for gaining
insight into model fit issues of the motivating palaoclimate example; see Sections \ref{sec:palaeo}, S-4, S-5 and S-6 for details.

%
For sufficiently large data sets, obtaining samples from all the leave-one-out posteriors $\{\pi(\cdot\mid X_{-i},Y);i=1,\ldots,n\}$
seems to be a daunting task. However, the IRMCMC methodology of \ctn{Bhatta06a} (see Section S-3 for an overview)
can be employed to generate samples from the inverse cross-validation posteriors in a very fast and efficient manner.   
Once samples from the leave-one-out posteriors 
are obtained, distributions of any discrepancy measures can be
trivially obtained using the samples. This is in sharp contrast with the methodology of \ctn{Bayarri00} 
(albeit their methodology is developed keeping the forward context in mind), since their proposal
requires re-computation for each discrepancy measure and hence is computationally burdensome. 

We finish this section by summarising the important differences between our approach and the approaches of
\ctn{Gelman96} and \ctn{Bayarri00} in Table \ref{table:differences}.

\section{Summary of further simulation studies illustrating our IRD approach} 
\label{sec:summary_simstudies}
In the supplement we consider five examples to illustrate our approach. Due to issues related to
space here we only consider a summary of the simulation studies and refer to the supplement for the details.

In Examples 1 and 2 we assume that given $x_1,\cdots,x_{10}$,
which are drawn randomly from $Uniform(1,2)$, the data $y_1,\ldots,y_{10}$ come from
$Geometric(p_i)$, where $p_i=1/(1+\theta x_i)$. 
We further assume that the data has been modeled as $Poisson(\theta x_i)$. A uniform improper
prior has been put on $\theta$, that is, $\pi(\theta)=1$; $\theta>0$.
In this case the two models are expected to agree closely when $\theta$ is small but increasing disagreement
is expected for increasingly large values of $\theta$. 
In Example 1 we consider the forward approach, where instead of constructing a reference distribution
of $T(\tilde X)$ we consider a reference distribution for $T(\tilde Y)$. Here $\tilde Y$ is defined analogously
as $\tilde X$. The forward approach is then compared and 
contrasted with our IRD approach of Example 2, where we assume $Uniform(1,2)$ prior on $x_i;i=1,\ldots,10$. 
The results of both the examples yielded the results expected---
that for small $\theta$, the incorrect
Poisson model has high chance of being accepted and high chance of rejection for high values of $\theta$. 
Interestingly, the forward approach displayed slightly greater power compared to our IRD approach.
This is to be expected since the forward approach only requires the probability model of $\tilde Y$ 
(which is the same as that of $Y$)
for computing $\pi(\tilde Y\mid X,Y)$, and the probability model of $Y$ is stronger than our weak prior
assumption on $\tilde X$ that we considered in the inverse approach. 

The undesirable relatively lower power of the IRD approach in the first two examples
prompted further investigation. That the power of our IRD approach can indeed
be improved with more informative priors on $\tilde X$ is the issue we demonstrate in Example 3.
In this example we assume that for $i=1,\cdots,10$, data $y_i$ 
come from the true model $Poisson(\theta x_i)$. 
The elements of $X=\{x_i;i=1,\cdots,10\}$ are drawn randomly from an exponential 
distribution with mean $\lambda$; this implies that the true prior for $\tilde X$ is given by
\[\pi(\tilde X)=\prod_{i=1}^{10}\pi(\tilde x_i)\]
where $\pi(\tilde x_i)$ are {\it iid} exponential with mean $\lambda$.
The parameter $\theta$ is selected randomly from the interval $(0,1)$.
Since the prior tends to be more and more flat for increasing $\lambda$, in Example 3 we
investigated if greater power results for an assumed non-informative prior on $\tilde X$ for large
values of $\lambda$ as opposed to smaller values of $\lambda$, assuming that the underlying Poisson model is known. 
The results of Example 3 confirm our anticipation.

In Example 4 we consider a variable selection problem, assuming the true model to be Poisson with mean 
$\theta=\theta_1 x_i +\theta_2 x^2_i$. 
We then assess which of the three cases: (a) $\theta=\theta_1 x_i$; (b) $\theta=\theta_1 x_i +\theta_2 x^2_i$
and (c) $\theta=\theta_1 x_i +\theta_2 x^2_i+ \theta_3 x^3_i$, is appropriate. 
Our IRD approach correctly identified the true model (b) most of the time.

In Example 5 we attempt to clarify the phenomenon of overfitting and that it can be detected
by our IRD approach. Here, given $\theta$ and $x_i$; $i=1,\cdots,10$ (drawn from
a uniform distribution), we assume that $y_i\sim Poisson(\theta x_i)$, but suppose that $y_i$ has been modeled as a Geometric
distribution with parameter $p_i=1/(1+\theta x_i)$. Here although the expected value of $y_i$ under
both the models is the same, the variance under the Geometric model 
is greater than in the Poisson case. Thus, for certain values of $\theta$ 
the Geometric model may overfit the data which actually comes from the Poisson model, and the discrepancy measure
in such a case may turn out to be too small, which would lead to acceptance of the Geometric model
unless our approach based on reference distribution is used. We present such a case
with $\theta=15$, where the discrepancy measure is small, apparently suggesting acceptance of the Geometric
model. But with respect to the reference distribution this measure is too small to lead to acceptance of the
wrong Geometric model, thus demonstrating a very desirable feature of our IRD approach.

We next consider application of our methodology to the motivating palaeoclimate example.

\section{Application of IRD approach to the motivating palaeoclimate example}
\label{sec:palaeo}

\ctn{Vasko00} reported a regular MCMC cross-validation exercise for a data set comprising
multivariate counts $y_i$ on $m = 52$ species of chironomid at $n = 62$ lakes (sites)
in Finland. The unidimensional $x_i$ denote mean July air temperature. As species respond
differently to summer temperature, the variation in the composition provides the analyst
with information on summer temperatures. This information is exploited to reconstruct
past climates from count data derived from fossils in the lake sediment; see \ctn{Korhola02}.

The cross-validation exercise was computationally challenging, requiring 62 separate regular MCMC exercises and involved
a parameter $\theta$ of dimension 3318. However, implementation of cross-validation by regular MCMC is not infeasible
in this case.
But the problem seems to be an ideal real life problem where the performance of IRMCMC
can be tested by making comparison with regular MCMC and complete details regarding this can be found
in \ctn{Bhattacharya04c}. 

In the case of \ctn{Vasko00}, our MCMC
implementation
took 16 hours.
In contrast, the IRMCMC implementation took
16 minutes
for
the initial run and
20 minutes for the remaining 61. Additionally, IRMCMC drew attention
to the bimodality of one of the posteriors, a point completely missed by the MCMC implementation.
For details, see \ctn{Bhatta06a}.
To proceed with the goodness of fit test, we first provide description of the underlying model.

\subsection{Model description}
\label{subsec:chironomid_model}

In \ctn{Vasko00}, the vector $y_{i}$ of counts at site $i$ followed the multinomial
distribution,
\begin{equation}
(y_i\mid y_{i+}, {\bf p}_i)\sim Multinomial(y_{i+}, {\bf p}_i).
\label{eq:multinomial}
\end{equation}
Here $y_i = (y_{i1},\cdots, y_{im})$,
$y_{i+} = \sum_{k = 1}^m y_{ik}$ and
${\bf p}_i$ is an (unobserved) vector of relative abundances $(p_{i1},\cdots,p_{im})$,
of dimensionality $(m-1) = 51$. We denote the multinomial likelihood as
\begin{equation}
L(y_i\mid y_{i+},{\bf p}_i) = \frac{(y_{i+})!}{\prod_{k=1}^{52} y_{ik}!}\prod_{k=1}^{52} p^{y_{ik}}_{ik}
\label{eq:multinomial_likelihood}
\end{equation}
The unobserved $\{{\bf p}_i; i = 1,\cdots, n\}$, thus provide $62\times 51$ parameters, even before
temperature $x_i$ is related to the relative abundances. \ctn{Vasko00} related these via
a Dirichlet model,
\begin{equation}
({\bf p}_i\mid x_i,\Psi_1,\cdots,\Psi_{52})\sim Dirichlet(\Lambda_i).
\label{eq:dirichlet}
\end{equation}
where the $k$th component $\lambda_{ik}$ of $\Lambda_i$ was modelled as

$\lambda_{ik} = \lambda(x_i,\Psi_k)$, for a simple function $\lambda$ of $x_i$ and of
$\Psi_k=(\alpha_k,\beta_k,\gamma_k)$, a 3-component parameter vector associated with the $k$th species. \ctn{Vasko00}
chose a simple unimodal ``response function" of these species specific parameters, given by
\begin{equation}
\lambda(x_i,\Psi_k)=\alpha_k\exp\left[-\left(\frac{x_i-\beta_k}{\gamma_k}\right)^2\right]
\label{eq:lambda}
\end{equation}
The mode, $\beta_k$, represents
the value of temperature at which the species $k$ is most abundant. Tolerance of the species
is denoted by $\gamma_k$ and $\alpha_k$ is a scaling factor. There are thus an
additional $3\times 52$ parameters, yielding $3318$ in total.
We write $\theta = \left\{{\bf p_1,\cdots,p_{62}},\Psi_1,\cdots,\Psi_{52}\right\}$.

As for the priors, \ctn{Vasko00} assume that $\alpha_k\sim Uniform(0.1,50)$, $\beta_k\sim Normal(11.19,1.57^2)$, $\gamma_k\sim Gamma(9,3)$
(that is, a Gamma distribution with mean 3 and variance 1) and $\tilde x_i\sim Normal(11.19,1.11^2)$.


\subsection{Results of assessment of model fit using the IRD approach}
\label{subsec:palaeo_results}

\begin{figure}
\centering
\leavevmode
\includegraphics[width=10cm,height=10cm]{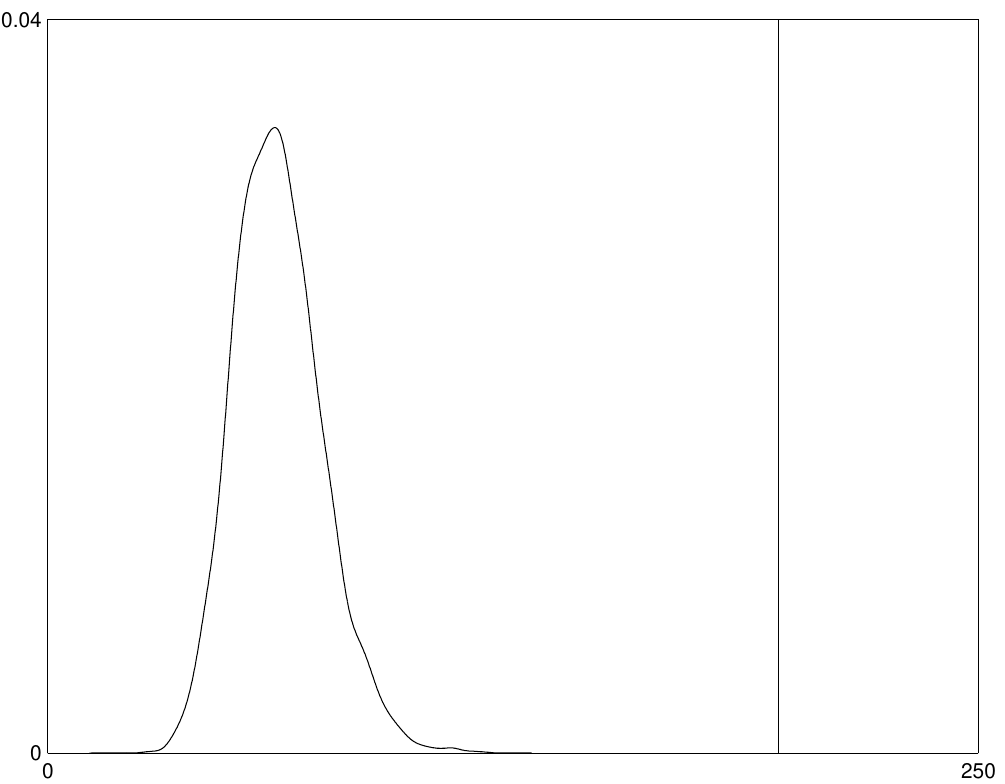}
\caption{Distribution of $D^{var}_1$; the vertical line is the observed
value, $D^{obs}_1$.
\label{fig:chironomid_chisq}
}
\end{figure}

Observed $T_1(X)$ and the posterior distribution of $T_1(\tilde X)$ are shown in Figure \ref{fig:chironomid_chisq}.
Note that $T_1(X)$ is located far from the mode of $T_1(\tilde X)$, indicating that the model does not fit the data. 
In fact, an application of the formal Bayesian hypothesis testing procedure
gives, for any sensible choice of $\epsilon$,
\[p=\pi\left(\left|\frac{T_1(\tilde X)-T_1(X)}
{\sqrt{V_{\pi}(T_1(\tilde X)\mid Y)}}\right|\leq\epsilon\mid Y\right)\approx 0.\]
This is a consequence of the
fact that many observed temperature values are far from the modes of the respective posterior density; 
see \ctn{Bhattacharya04c}. In fact,
it has been found that more than 40\% of the observed data lie outside
the 95\% highest posterior density credible regions, suggesting poor fit of the model to the data.
We anticipated that the reason for this lack of fit is that the assumed unimodal model used to describe $\lambda_{ik}$ in (\ref{eq:lambda}) is
inappropriate. Indeed, it has been argued in the palaeoclimate literature that species can have multiple climate preferences, in which case
the unimodal model is inappropriate.
\ctn{Bhatta06b} used another modelling approach where, rather than unimodal functions,
the response functions $\lambda_{ik}$ were modelled as mixtures of normal densities;
the number of components being unknown.
He viewed the parameters associated with each component of the mixture as samples arisen from
the Dirichlet process (see, for example, \ctn{Ferguson73}, \ctn{Ferguson83}, \ctn{Escobar95}).
This way of modelling automatically induces a prior on the number of components; see
\ctn{Antoniak74}.  This approach to modelling the response surfaces improved the model fit, although it is yet to be completely
satisfactory. In Section S-5 we describe this in detail.

\section{Conclusions}
\label{sec:conclusions}


The IRD approach is simple and we have attempted to provide clear cut guidelines when to accept or reject the model
in question. It also seems to have very general applicability.
A key point of our proposal is that it does not recommend 
acceptance or rejection of a model by noting the magnitude of an observed discrepancy measure alone. 
%

One important point
to note is that there are no parameterisation problems in our approach with reference distributions since
all parameters other than $\tilde X$ are integrated out.
No asymptotic theory, of any sort,
is needed to make this approach work. 
Importantly, the data is not used twice and P-values have been replaced with Bayesian credible regions.
In particular, the latter point makes our approach ``more Bayesian" compared to the other available approaches. 

Simulation from the cross-validaton posteriors $\pi(x\mid X_{-i},Y)$
for each $i$, needed to compute the  
reference distribution corresponding discrepancy measure, appears very demanding at the first sight, particularly if there are a large number 
of cases. However, IRMCMC is a method that seems to
be highly suitable for computing the pairs cheaply and efficiently. 
We recommend IRMCMC
for the computational needs of this model assessment proposal.

All said, however, there is certainly scope for further investigations. 
In fact, our aims regarding this paper is quite modest --- to indicate potential advantages as well as
to shed light on issues involved in our proposal that need future attention.
For instance, the issue regarding the prior on $\tilde X$ deserves
more careful attention. We believe, that for appropriate informative prior it is possible 
to overcome the slight deficiency
of the power that our approach seems to currently exhibit. 
One important issue that we ducked in this paper concerns questions regarding appropriate discrepancy measures. 
It will be interesting to address optimality properties of discrepancy measures.
There may also be questions regarding the reliability of the IRD approach if the posterior distribution of $\tilde x_i$, for some, or all $i$, are multimodal. 
However, in such cases, there exist choices
of $T$ for which the posterior distribution of $T(\tilde X)$ will be unimodal; see, for example,
\ctn{Baker30}. 
In the assessment of Vasko's model, one $\tilde x_i$ became bimodal,
and in our model in \ctn{Haslett06}, most $\tilde x_i; 1=1,\ldots,7815$ were multimodal; for details,
see \ctn{Bhattacharya04b}, \ctn{Bhattacharya04c}.
But the discrepancy measure $T_1(\tilde X)$ was unimodal in all cases.


Another topic for future research is to systematically address
the question of the efficiency of our proposal when the number of covariates is allowed to be very large. Although the general methodology presented in this paper
will certainly remain valid for multi-dimensional covariates, we anticipate that it may be slightly difficult to devise appropriate overall discrepancy measures
of goodness of fit. It is worthwhile to note in this connection that 
\ctn{Bhattacharya04b} addressed goodness of fit of the complicated
palaeoclimate model of \ctn{Haslett06}, where there are two covariates instead of one; see also \ctn{Bhattacharya04c} for more detail. In the above-mentioned research
two separate discrepancy measures were constructed instead of a single overall measure of fit. The final conclusions regarding goodness of fit of the model, however, 
were consistent with respect to the two independent discrepancy measures.
More recently, \ctn{Sabya13} proposed a new Bayesian palaeoclimate model and were able to confirm goodness-of-fit of their model to the chironomid data of \ctn{Vasko00}
and the pollen data of \ctn{Haslett06}, using the IRD method and several discrepancy measures, all of which led to the same conclusion of good fit.
We also remind the reader that our proposal 
is ideally suited for model assessment in inverse regression problems. However, this
seems to have quite good potential in assessing Bayesian model fit in general.
We look forward to
providing a detailed separate paper on issues regarding application of our methodology
to problems other than inverse regression.
%

\section*{Acknowledgments}
We are grateful to an anonymous referee for providing very encouraging feedback 
and very useful suggestions on
an earlier version of this manuscript which resulted in much better presentation of our ideas. 
We also acknowledge a very useful ``e-discussion" with Professor Jayanta Ghosh. 
The author also thanks Kari Vasko for providing the chironomid
data set and Professor Jim Berger for commenting on this paper. 
Part of the work was done when the author was pursuing 
his PhD at Trinity College Dublin; the author had some useful
discussions with Professor John Haslett.




\begin{table}
\caption{Comparison of reference distribution approaches\label{table:differences}}
\begin{center}
\begin{tabular}{|c|c|c|}\hline
\multicolumn{1}{|c|}{\rule[-5mm]{0mm}{10mm}\mbox{{\bf IRD}}}
& \multicolumn{1}{|c|}{\rule[-5mm]{0mm}{10mm}\mbox{{\bf Bayarri and Berger}}}
& \multicolumn{1}{|c|}{\rule[-5mm]{0mm}{10mm}\mbox{{\bf Gelman et al.}}}\\
\hline
${\rule[-3mm]{0mm}{6mm} \mbox{Fully Bayesian approach}}$ & $\mbox{Not fully Bayesian}$ & $\mbox{Not fully Bayesian}$\\
\hline
${\rule[-1.5mm]{0mm}{6mm} \mbox{Uses $\pi(\tilde X\mid Y)$}}$ & $\mbox{Uses a modified version of $\pi(\tilde Y\mid X,Y)$}$ & $\mbox{Uses $\pi(\tilde Y\mid X,Y)$}$\\
${\rule[-1.5mm]{0mm}{6mm} \mbox{as reference distribution}}$ & $\mbox{as reference distribution}$ & $\mbox{as reference distribution}$\\
\hline
${\rule[-3mm]{0mm}{6mm}\mbox{Measure independent of $Y$}}$ & $\mbox{Depends on $Y$}$ & $\mbox{Depends on $Y$}$\\
\hline
${\rule[-3mm]{0mm}{6mm}\mbox{Measure independent of $\theta$}}$ & $\mbox{May depend on $\theta$}$ & $\mbox{May depend on $\theta$}$\\
\hline
${\rule[-3mm]{0mm}{6mm}\mbox{Avoids double use of data}}$ & $\mbox{Asymptotically avoids double use}$ 
& $\mbox{Uses data twice}$\\
\hline
${\rule[-1.5mm]{0mm}{6mm}\mbox{Uses credible sets,}}$ & $\mbox{Uses P-values}$ & $\mbox{May use credible sets}$\\
${\rule[-1.5mm]{0mm}{6mm}\mbox{not P-values}}$ &  & $\mbox{but directly related to P-values}$\\
\hline
${\rule[-3mm]{0mm}{6mm}\mbox{Has calibration property}}$ & $\mbox{Asymptotically has calibration property}$ & $\mbox{No
calibration property}$\\
\hline

${\rule[-3mm]{0mm}{6mm}\mbox{Computation easy}}$ & $\mbox{Computation hard}$ & $\mbox{Computation easy}$\\

\hline
\end{tabular}
\end{center}
\end{table}




%
\newpage

\renewcommand\thefigure{S-\arabic{figure}}
\renewcommand\thetable{S-\arabic{table}}
\renewcommand\thesection{S-\arabic{section}}

\setcounter{section}{0}
\setcounter{figure}{0}
\setcounter{table}{0}

\begin{center}
	{\bf \LARGE Supplementary Material}
\end{center}

\section{Discrepancy measures for model assessment in inverse problems}
\label{sec:discrepancies}

Using an inverse cross-validation approach, we first simulate, for each $i=1,\cdots,n$, $N$ realisations 
from the distribution $\pi(\tilde x_i\mid X_{-i},Y)$. Let the simulated values be denoted by 
$\{\tilde x^{(1)}_i,\cdots,\tilde x^{(N)}_i\}$.
The simulation can be carried out very efficiently by using a methodology proposed by \ctn{Bhatta06a};
we discuss this briefly in Section \ref{sec:computation}. 
Some examples of $T(X)$ are as follows:
\begin{eqnarray}
T_1(X) &=& \sum_{i=1}^n\frac{(x_i - E_{\pi}(\tilde x_i))^2}{V_{\pi}(\tilde x_i)}
\label{eq:discrepancy1}\\
T_2(X) &=& \sum_{i=1}^n\frac{|x_i - E_{\pi}(\tilde x_i)|}{\sqrt{V_{\pi}(\tilde x_i)}}
\label{eq:discrepancy2}\\
T_3(X) &=& \max_{1\leq i\leq n}\left\{\frac{|x_i - E_{\pi}(\tilde x_i)|}{\sqrt{V_{\pi}(\tilde x_i)}}\right\}
\label{eq:discrepancy3}\\
T_4(X) &=& x_i
\label{eq:discrepancy5}
\end{eqnarray}

We make no argument on the merits and demerits of the above discrepancy measures. 
However, note that while measures $T_1(X)$, $T_2(X)$ and $T_3(X)$ provide summaries of distances
between the observed values $x_i$ and the corresponding summaries of the leave-one-out posteriors $\pi(\cdot\mid X_{-i},Y)$,
the measure $T_4(X)$ is just the observed value for case $i$ and thus is different from all other measures in
the sense that it is not an overall measure of fit. Rather it provides insight
specifically into the case $i$. For example, it can be used to check whether or not $x_i$ is an outlier with respect
to the underlying model. In this context we note that there may exist measures corresponding to which no
reference distribution may be easily available. For instance, a measure $T_5(X)$ may be defined as the number of
$x_i$ that fall within the $100(1-\alpha)$\% credible region of the corresponding leave-one-out posteriors. In this case
there seems to exist no easily computable reference distribution.


\section{Discussion regarding priors on $(\tilde X,\theta)$}
\label{sec:priors}

It is important to have some discussion on the choice of priors on $\tilde X$ and $\theta$. 
Since $\theta$ is a set of model parameters, the issue of choice of the prior on $\theta$ is generic. In the absence
of any information, which is usually the case, it is natural to put a somewhat vague
prior (usually non-informative) on the model parameters $\theta$, hoping that the true value is supported by the prior. 
In our illustrations, we use
non-informative (improper) priors on $\theta$. 

The issues regarding the prior on $\tilde X$ are more interesting. Note that the true values $X$ are known, so it is tempting to 
put an overly strong prior on $\tilde X$ which assigns all mass to the true values $X$. However, this choice of prior is certainly
inappropriate for assessing model fit, since irrespective of the suitability of the model to the data, the posterior of $T(\tilde X)$ will be a point probability mass
at $T(X)$, thus reflecting only the prior aspect. For proper model checking it is necessary to make the prior parameter space of $\tilde X$ as large as possible, so that
all possible values of the covariates are explored. One can then observe, whether or not the observed covariates get high density {\it a posteriori}. 

The preceding discussion
seems to suggest a non-informative prior for $\tilde X$. 
However, if prior information of the covariates is available, then there is no reason not to use the information to construct an appropriate
prior for $\tilde X$. In fact, Example 3 of Section \ref{sec:illustration} demonstrates that when prior information about $\tilde X$ is available, then it is
less efficient to put a non-informative prior on $\tilde X$. 
In the palaeoclimate study reported in \ctn{Haslett06} prior information on the past climates were available, which were used to reconstruct past climates from fossil
pollen data. 
In the palaeoclimate example in Section 7 of \ctn{Bhattacharya12} we use available 
prior information on the unknown covariates to implement our proposed model assessment idea. In that problem it is assumed that the components of $\tilde X$ are {\it a priori} $iid$.
We remark here that the priors for both modern and past climates are obtained from experts before observing the data (modern or fossil). When the past is not too far from the present,
one can use the same prior for both modern and past climates. For general problems, however, such prior information
will not be available. In such cases one possibility is to estimate some features of the prior distribution using empirical Bayes analysis. In fact, the latter procedure, which
uses data to reliably estimate features of the prior distribution, has received wide attention in the Bayesian statistical literature. For details on this procedure see \ctn{Berger85}; see also \ctn{Carlin00}.  Going by the principles
of empirical Bayes methods, it is not unreasonable to estimate at least some features, say, moments of the prior distribution on $\tilde X$ based on observed covariates $X$. It is important to remind the reader in this context that
strictly speaking, $X$ is not the data, since unlike in the case of $Y$,  there is no probability model associated with $X$. Only $Y$, which has a probability model, given the covariates $X$, 
should be strictly regarded as the data. Hence we argue that estimation of some
features of the prior on $\tilde X$ using observed $X$ is a reliable procedure; it is neither non-Bayesian, nor does it lead to double use of the data.
As an aside, and as a possible topic for future research, we note that it may also be advisable to check robustness of the results on model assessment with respect to several plausible priors on $\tilde X$, including the
one obtained by empirical Bayes analysis. 
In the palaeoclimate study reported in \ctn{Haslett06} prior information on the past climates were available, which was used to reconstruct past climates from fossil
pollen data. However, \ctn{Haslett06} also performed limited sensitivity analysis; for complete details, see \ctn{Bhatta06c}. In this research, however, we do not discuss sensitivity analysis.

\section{Computation of inverse leave-one-out posteriors}
\label{sec:computation}

Sampling from the cross-validation posteriors $\pi(\tilde x_i\mid X_{-i},Y);i=1,\ldots,n$ seems to be
very demanding at the first glance, since, in principle, $n$ many computer-intensive runs of regular MCMC,
which we call $n$-fold regular MCMC, are necessary. 
\ctn{Bhatta06a} show that the approach proposed by \ctn{Gelfand92}, \ctn{Gelfand96} which is based on 
importance sampling (see, for example, \ctn{Geweke89}) in the 
context of forward problems is inapplicable to inverse problems.  
However, a novel methodology proposed 
by \ctn{Bhatta06a} seems
to be very promising in this regard. The above authors refer to the methodology
as Importance Resampling MCMC (IRMCMC). The key idea is to leave out case $i^*$, to sample by regular MCMC
realizations of $(\tilde x_{i^*},\theta)$, given $X_{-i^*},Y$, and to draw a subsample of $\theta$ values using
appropriately constructed importance weights. 

Given each re-sampled $\theta$, MCMC may be used to realise
$\tilde x_i$ from the conditional distribution of $\pi(\cdot\mid y_i,\theta)$. In particular, MCMC needs to be carefully
implemented once, to a selected case $i^*$, generating realisations of $(\tilde x_{i^*},\theta)$. For all cases other than $i^*$
the resultant sample of $\theta$ values may be re-used using importance resampling (IR) (see, for example, \ctn{Rubin88}).
In fact, the proposal of \ctn{Bhatta06a} is equivalent to resampling both $(\tilde x_i,\theta)$ using importance
resampling but subsequently ignoring $\tilde x_i$, retaining $\theta$ only. Critically, for each such $\theta$, sampling
would be done from the low-dimensional $\pi(\cdot\mid y_i,\theta)$, for constant $\theta$, typically by MCMC.
The latter exercise is very fast. Choice of $i^*$ has been discussed in detail by \ctn{Bhatta06a}; in particular,
they show that it is easy to choose $i^*$ appropriately.

The proposed procedure of \ctn{Bhatta06a} can be stated in the following manner.
\\[2mm]
1.  Choose an initial case $i^*$. Use $\pi(\tilde x_i,\theta\mid X_{-i^*}, Y)$ as the importance
sampling density.
\\[2mm]
2.  From this density, sample values
$(\tilde x^{(\ell)},\theta^{(\ell)}); \ell = 1,\cdots,N$, for large $N$.
Typically, regular MCMC will be used for sampling.
\\[2mm]
3.  For $i\in\{1,\cdots,i^*-1, i^*+1,\cdots,n\}$ do
\begin{itemize}
\item[a.] For each sample value $(\tilde x^{(\ell)},\theta^{(\ell)})$, compute importance weights
$w^{(\ell)}_{i^*,i}=w_{i^*,i}(\tilde x^{(\ell)},\theta^{(\ell)})$, where the importance weight function is
given by
\begin{equation}
w_{i^*,i}(\tilde x_i,\theta) =\frac{\pi(\tilde x_{i^*},\theta\mid X_{-i},Y)}{\pi(\tilde x_{i^*},\theta\mid X_{-i^*},Y)}
\propto\frac{L(Y,X_{-i},\tilde x_i,\theta)}{L(Y,X_{-i^*},\tilde x_i,\theta)}
=\frac{f(y_{i^*}\mid x_{i^*},\theta)f(y_{i}\mid \tilde x_{i^*},\theta)}{f(y_{i^*}\mid \tilde x_{i^*},\theta)f(y_{i}\mid x_{i},\theta)}.
\label{eq:better_weights}
\end{equation}
In the above, $L(Y,X,\theta)$ is the
likelihood of the observed data under the model.

\item[b.] For $k\in\{1,\cdots,K\}$
\begin{enumerate}
\item[(i)] Sample $\tilde\theta^{(k)}$ from $\theta^{(1)},\cdots,\theta^{(N)}$
where the probability of sampling $\theta^{(\ell)}$ is proportional to
$w^{(\ell)}_{i^*,i}$.

\item[(ii)] For {\it fixed} $\theta=\tilde\theta^{(k)}$, draw $M$ times from $\pi(\tilde x_i\mid y_i,\tilde\theta^{(k)})$.
Note that in general it is not
easy to sample from $\pi(\tilde x_i\mid y_i,\tilde\theta^{(k)})$, even  if $\tilde x_i$ is univariate, and
we recommend MCMC for generality. 
\end{enumerate}

\item[c.] Store the $K\times M$ draws of $\tilde x_i$ as
$\tilde x_i^{(1)},\ldots,\tilde x_i^{(KM)}$.

\end{itemize}

The key idea in the above proposal
is the use of $\pi(\tilde x_i,\theta\mid X_{-i^*},Y)$ as the importance sampling
density, for some particular $i^*$. 
\ctn{Bhatta06a} demonstrate that it is easy to choose an appropriate $i^*$. 

It is shown by \ctn{Bhatta06a} that IRMCMC is MCMC
with a special proposal kernel. They demonstrate that compared to $n$-fold regular MCMC, IRMCMC is many times
faster than regular MCMC and mixes as least as good as regular MCMC.
That IR yields reliable approximation in this cross-validation proposal is clear, since IR is used only to sample $\theta$ and the importance sampling
density $\pi(\theta\mid X_{-i^*},Y)$ is a good approximation to $\pi(\theta\mid X_{-i},Y)$ for any $i$. It is important to note that the posterior $\pi(\tilde x_{i^*}\mid X_{-i^*},Y)$
is generally not a good approximation for $\pi(\tilde x_i\mid X_{-i},Y)$; in fact, they usually have disjoint supports. So, very reasonably, we have avoided IR to sample from
the required $\pi(\tilde x_i\mid X_{-i},Y)$, using instead resampled values of $\theta$ to sample, via regular MCMC, from $\pi(\tilde x_i\mid y_i,\theta)$.
An important technical question is whether IR should be used with or
without replacement. Although most of the references to IR in the
literature recommend IR with replacement (see, for example, \ctn{Gelfand92}, \ctn{Newton94}, \ctn{OHagan04}), \ctn{Gelman95}, \ctn{Stern00} recommend IR without replacement. They
argue that sampling without replacement can provide protection
against highly variable importance weights. In fact, \ctn{Skare03}
formally prove a theorem that with respect to the total variation norm, IR without replacement is better than
IR with replacement. Hence \ctn{Bhatta06a}
recommend IR without replacement. \ctn{Bhattacharya04c}
provides further details in this context including a comparison of
IR with/without replacement.

\section{Illustration of inverse model assessment with the reference distribution approach}
\label{sec:illustration}

It has been argued that 
the reference distribution approach in inverse problems, which is the main contribution 
in this work, has some desirable properties and that the computational challenge
involved may be overcome by IRMCMC.
We now illustrate the approach by applying it on various problems involving
repeated computer-simulated data and mainly noting
the percentage of times it gives the correct answer. 
However, we acknowledge
that since we obtain only point estimates of true percentages, our evaluation procedure
may not be completely adequate.

In the following illustrations we emphasize that experimental evaluation sheds more 
light on the particular choice of discrepancy
measure. Even on any particular choice of discrepancy measure experimental replications can shed limited light.
But since here we are concerned with simulation studies,
where the true models and their properties are completely known, we may suppose that
the point estimates provide useful evidence on the general performance of our
approach based on reference distributions. Besides, we provide other relevant experimental details to supplement
the inadequacy of the point estimates.

In none of our examples do we claim optimality of any particular discrepancy measure.
Throughout all illustrations the results based on the discrepancy measure
$T_1(X)$ and the corresponding reference distribution $T_1(\tilde X)$ will be presented.
In the examples, we consider that the model fits the data if $T_1(X)$ falls within approximately 97\% credible region
of $T_1(\tilde X)$ (here 97\% does not have any special significance, but we chose this
merely because we found that, in the experiments the percentage of times the correct answer is obtained
is often close to 97\% if 97\% credible regions of $T_1(\tilde X)$ are chosen! We could have certainly
chosen $100(1-\alpha)\%$ credible region of $T_1(\tilde X)$ for any $0<\alpha<1$).
\\[3mm]
{\bf Example 1: Forward regression}
\\[3mm]
Our first example concerns a forward problem. But we consider this forward regression problem mainly to 
contrast it with later examples on inverse regression problems. 
In this example, the data actually comes from
a Geometric distribution but has been modeled in reality as involving
the Poisson distribution. In other words, given $x_1,\cdots,x_{10}$,
which are drawn randomly from $Uniform(1,2)$, data $y_i\sim Geometric(p_i)$,
where $p_i=1/(1+\theta x_i)$. 
It is assumed, for purposes of illustration,
that the data has been modeled as $Poisson(\theta x_i)$. A uniform improper
prior has been put on $\theta$, that is, $\pi(\theta)=1$; $\theta>0$.
%
%
%
Note that, had $y_i$ been $Poisson(\theta x_i)$, then $E(y_i)=\theta x_i=Var(y_i)$.
But for the Geometric case, $E(y_i)=\theta x_i$ but $Var(y_i)=E(y_i)(1+\theta x_i)$. 
In this example $x_i\in (1,2)$ and $\theta>0$. Since $x_i$ are bounded, for $\theta$ close to zero $Var(y_i)\approx E(y_i)$ 
and we can expect Poisson and Geometric distributions to agree. 
However, if $\theta$ is large, then $Var(y_i)>> E(y_i)$ and the two distributions are expected
to disagree.
%
%
%
%
We considered 1000 simulations from the Geometric distributions with the above set-up 
with different values of $\theta$ and applied
our methodology in each case to assess the goodness-of-fit
of the Poisson model to the Geometric data. Subsequently the true model has also been applied
on the data to contrast with the fit achieved by the Poisson model. The results
are given in Table \ref{table:fwd_pg}. For example, for $\theta=0.1$ and the true model is Geometric,
the erroneous Poisson model is accepted 97\% times (false positive) and the true Geometric model is rejected 0.3\% times
(false negative).
In Example 2 we will contrast this with
the inverse case.



Observe that, as $\theta$ increases, the Poisson model agrees less and less with the Geometric model. 
This is because the mean and the variance of the Geometric distribution drift apart as $\theta$ increases. 
In fact,
the percentages of agreement by the Poisson model decrease quite fast. It will be pointed out
that in the inverse case that the decrease is relatively slow in comparison. 
Note that when the Geometric model is applied to the data it fits the data very well in all the cases.
This is to be expected since it is the true model. In the inverse case it will be seen that the
percentages of agreement by the Geometric model are comparatively slightly less. It will be argued that
at the cost of other theoretical and computational advantages, the inverse model checking approach may have slightly less power compared to the forward approach.
\\[3mm]
{\bf Example 2: Inverse regression}
\\[3mm]
In the first example we considered a problem involving the Geometric distribution
as the true model but modeled as Poisson distribution. There assessment of the model
fit used pairs $\{y_i,\pi(\tilde y_i\mid X,Y_{-i})\}$. 
In this example we consider the same problem but now we focus on the pairs
$\{x_i,\pi(\tilde x_i\mid X_{-i},Y)\}$ instead. We remind the reader that herein lies our interest.
In contrast to the previous forward example, here we need to put a prior on $\tilde x_i$ (in addition
to the prior on $\theta$, which we assume the same as in the previous example). We put the
correct prior on $\tilde x_i$; that is $\tilde x_i\sim Uniform(1,2)$ (recall that in Example 1 $\tilde x_i$ has been drawn randomly
from $Uniform(1,2)$).

We provide in Table \ref{table:inv_pg} 
abridged results of varying
the parameter $\theta$. The table clearly shows that Poisson disagrees more and more with the
Geometric model as $\theta$ increases, and thus the difference between mean and variance of the Geometric model, increases. 
In other words, the percentage of false positives decreases fast as $\theta$ increases. On the other hand, the
percentage of false negatives do not show any appreciable change with $\theta$.
It is important to note that, in this example, we have used the true prior for $\tilde X$, but as mentioned before this has no implication 
on the model adequacy test of our proposed inverse approach; it does not accept the model when it is false. In other words, even though the prior on $\tilde X$ is assumed to be correct, our proposal correctly rejected the model
corresponding to the incorrect probability distribution of $Y$ and correctly accepted the model whenever the probability distribution of $Y$ is sufficiently close to the true
probability model.
But here one must note the contrast between Table \ref{table:fwd_pg}, of the forward case, and 
Table \ref{table:inv_pg}, corresponding to the inverse case. In the former
table the percentage of disagreement of the Poisson model with the Geometric data increases slightly faster
than in the table corresponding to the inverse case. Also, the percentages of agreement of the Geometric
model with the true Geometric data is slightly higher in the forward case. These observations indicate that the power of the test
with the inverse approach is slightly less than that with the forward approach. This is because, with the inverse approach, a {\it prior}
on $\tilde X$ is used, which is generally weaker than the probability model of $Y$, which is used to compute $\pi(\tilde Y\mid X,Y)$ in the forward
approach. However, slightly less power is not to be interpreted as a major drawback of our approach. Certainly, the results with the inverse approach very clearly
assert the reliability of our proposal, in spite of slightly less power. Moreover, we have already discussed in detail in our main manuscript \ctn{Bhattacharya12} that the inverse approach has a solid theoretical
framework and nice computational properties as compared to the other available approaches. We also believe that with informative priors on $\tilde X$, constructed from
observed $X$ using empirical Bayes analysis, will make up for the slight loss of power.




We now introduce an example to check the prior assumption on $\tilde X$ in an inverse problem. In particular we demonstrate that, when prior information about $\tilde X$
is available, then using a non-informative prior for $\tilde X$ is inefficient for model-checking purpose.
\\[5mm]
{\bf Example 3: Inverse regression -- implications of prior assumptions on $\tilde X$}
\\[3mm]
For $i=1,\cdots,10$, data $y_i$ 
come from the true model $Poisson(\theta x_i)$. 
Data $X=\{x_i;i=1,\cdots,10\}$ are drawn randomly from an exponential 
distribution with mean $\lambda$. In other words, the true prior for $\tilde X$ is given by
\[\pi(\tilde X)=\prod_{i=1}^{10}\pi(\tilde x_i)\]
where $\pi(\tilde x_i)$ are {\it iid} exponential with mean $\lambda$.
The parameter $\theta$ is selected randomly from the interval $(0,1)$.

Given the above set up we now assume that it is known to us that $y_i\sim Poisson(\theta x_i)$, but
that the prior distribution of $\tilde X$ is unknown. We test whether a uniform improper prior is appropriate for $\tilde X$. 

We evaluate our approach with several different
true values of $\lambda$. Note that for an exponential distribution with mean $\lambda$, the variance
is $\lambda^2$; since uniform improper priors can be said to have infinite variance, we can expect 
the fitted model to agree with the true model when $\lambda$ is large and disagree when $\lambda$ is small.
We summarise our findings in Table \ref{table:inverse_model_fit}. Very clearly, the results are in keeping with our expectations.
Unless the true prior for $\tilde X$ is reasonably flat, the assumed uniform improper prior is not very appropriate. In fact, the conclusions
drawn from this example are in agreement with the power issue discussed in the first two examples. From the current example it is clear that
a properly elicited informative prior, which can be thought of as a good representative of the true prior, can improve the power of the inverse approach. 
We remark here that a prior for $\tilde X$ estimated using observed $X$ and principles of empirical Bayes analysis is likely to approximate the true prior
very accurately and hence
will be far more appropriate than the uniform improper prior used. It has already been argued why using observed $X$
to construct the prior for $\tilde X$ makes sense.
\\[3mm]
{\bf Example 4: Variable selection}
\\[3mm]
In addition to the above three examples, we have also conducted a variable selection
study, assuming the true model to be Poisson with mean $\theta=\theta_1 x_i +\theta_2 x^2_i$.
Here the true values of $\theta_1$ and $\theta_2$ are 0.5 and the $x_i$ were drawn randomly from
$Uniform(0,10)$. As in the previous examples, here also we will present results based on $T_1(X)$
and $T_1(\tilde X)$ only.

Given the above set up, we consider three cases: (a) $\theta=\theta_1 x_i$; (b) $\theta=\theta_1 x_i +\theta_2 x^2_i$
and (c) $\theta=\theta_1 x_i +\theta_2 x^2_i+ \theta_3 x^3_i$. Clearly, except (b), others are incorrect.



For each of the three models (a), (b) and (c), with the simulation procedure repeated 1000 times, 
we implement our approach based on $T_1(X)$ and $T_1(\tilde X)$ by simulating from the leave-one-out 
posteriors $\pi(x\mid X_{-i},Y)$, corresponding to uniform priors for all variables. Case (b) was
adjudged the correct model 95\% times, cases (a) and (c) agreed with the true model 39\% and 84\% times
respectively. 

It is not at all surprising that (c) turns out to be far better than (a); this is because
(a) wrongly assumes that $\theta_2=0$ but (c) does not neglect the quadratic term. In fact, in addition, (c) considers
an extra cubic term. Noting that the true model (b) can be written as  
$\theta=\theta_1 x_i +\theta_2 x^2_i+0\times x^3$, the true value of $\theta_3$ in (c) 
can be said to to be zero. 
We remark that obtaining realisations from the leave-one-out posteriors is simple
in the simple examples provided. However, this is certainly not a simple exercise
in the case of the real examples reported in Section 7 of our main manuscript, Section \ref{sec:dp},
and Section \ref{sec:pollen}. 
In those cases
IRMCMC is clearly necessary.

\begin{figure}
\centering
\leavevmode
\includegraphics[width=10cm,height=10cm]{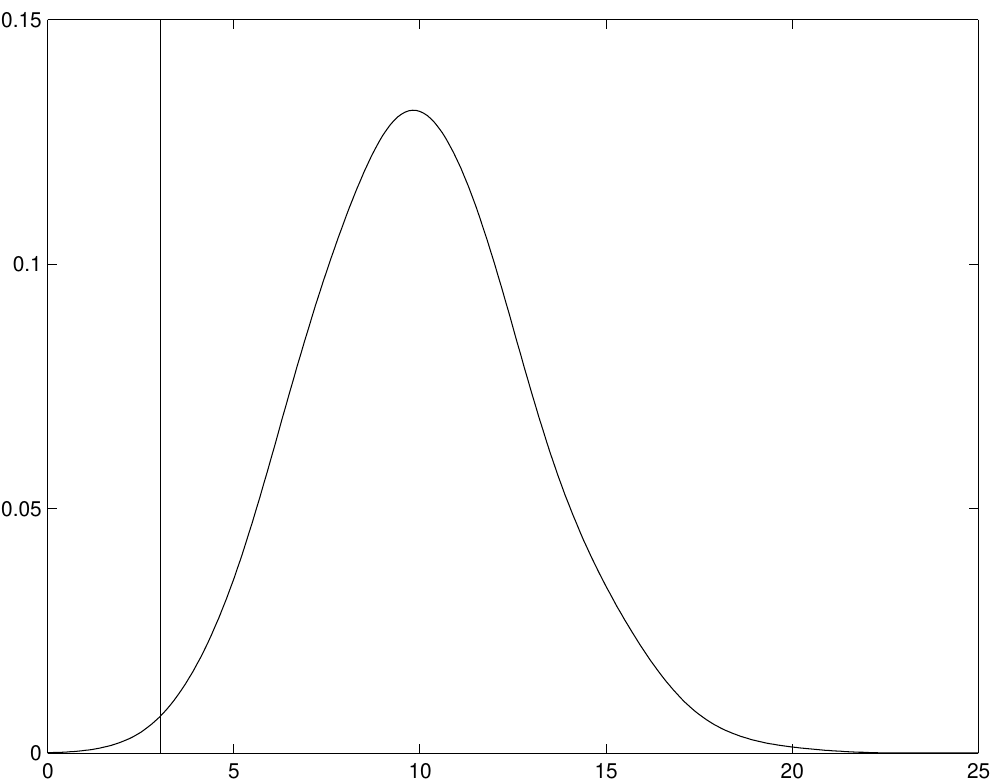}
\caption{Demonstration of overfitted situation in an inverse problem involving Poisson and Geometric model. 
Here considering solely the observed discrepancy measure
(denoted by the vertical line) wrongly leads to acceptance of the overfitted model; considering it with respect to the reference
distribution leads to the correct decision (that is rejection of the model).
\label{fig:pg_overfitted}
}
\end{figure}

We next discuss the use of reference distributions in detecting overfitting in models.
In particular we demonstrate that even when the observed discrepancy measure is small,
this does not necessarily lead to acceptance of the model in question.
In such cases, basing decisions solely
on the smallness of the magnitudes of the observed
discrepancy measures may be quite misleading.
Below we illustrate this with an example. 
\\[3mm]
{\bf Example 5: Overfit in inverse regression}
\\[3mm]
Given $\theta$ and $x_i$; $i=1,\cdots,10$ which arise from
a uniform distribution, 
suppose that $y_i\sim Poisson(\theta x_i)$. But suppose that $y_i$ has been modeled as a Geometric
distribution with parameter $p_i=1/(1+\theta x_i)$. Here although the expected value of $y_i$ under
both the models is the same, given by $\theta x_i$, the variance under the Geometric model given by $\theta x_i(1+\theta x_i)$
is greater than that in the Poisson case, where it is given by $\theta x_i$. Thus, for certain values of $\theta$ 
the Geometric model may overfit the data which actually comes from the Poisson model. Figure \ref{fig:pg_overfitted}
presents a case with $\theta=15$.  
%
%
%
%
%
In this case the discrepancy measure is too small with respect to the reference distribution. Thus the Geometric model is to be
considered a poor fit to the observed Poisson data. \ctn{Bhattacharya04c} (see chapters 7 and 9) discusses two real cases of overfitting. 

\section{Improving palaeoclimate model by modelling response surface as a mixture of Gaussian curves}
\label{sec:dp}


In Section 7.2 of \ctn{Bhattacharya12} it is shown that the model of \ctn{Vasko00} does not fit the data. Further investigation 
using exploratory data analysis
suggested that the unimodal model to relate species to environment may not be appropriate. In fact, in the palaeoclimate literature this unimodal model
has been criticised on the ground that each species may have multiple climate preferences. Also, some species may represent an entire genus consisting of
many sub-species, where each sub-species may have different climate preferences. So, even if the response curve for each sub-species is
unimodal, the response curve for the genus of species is certainly not unimodal. For general discussion on this, see \ctn{Haslett06}. 

In order to obtain an improved version of the model of \ctn{Vasko00} that takes into account the multimodal nature of response curves,  \ctn{Bhatta06b} introduced a novel approach based
on Dirichlet process to model a very flexible class of multimodal
models to relate species to environment. We begin to describe his approach by defining, as
an analogue of (17) of \ctn{Bhattacharya12}, the following

\begin{equation} 
 {\lambda^*}_{ik}=\sum_{j=1}^{r_k}R_{kj}\frac{1}{\gamma_{kj}\sqrt{2\pi}}{\exp}
 {\left[-\left(\frac{x_i-{\beta}_{kj}}{{\gamma}_{kj}}
 \right)^2
 \right]}
 \label{eq:lambda2}
 \end{equation}
where for $k=1,\ldots,m$, $r_k$ is a discrete random variable
taking values between 1 and $R_{k+}$ (\textit{both inclusive}), where 
given {\it fixed} $r_k$, $R_{k+}=\sum_{j=1}^{r_k} R_{kj}$. 

The above implies that the response function for the $k^{th}$ species given by (\ref{eq:lambda2}) is a mixture of Gaussian densities;
the number of mixture components, denoted by $r_k$, being unknown and hence regarded as a random variable. Certainly, it also includes the unimodal
model as a special case, when the components of the mixture are all equal. Thus the response function
is a multimodal function, with the number of modes (and indeed the magnitudes of the modes, $\beta_k$, and the scales, $\gamma_k$) being unknown.

Equation (\ref{eq:lambda2}) can be re-written as
\begin{equation}
 {\lambda^*}_{ik}=\sum_{j=1}^{R_{k+}}\frac{1}{\gamma_{kj}\sqrt{2\pi}}{\exp}
 {\left[-\left(\frac{x_i-{\beta}_{kj}}{{\gamma}_{kj}}
 \right)^2
 \right]}
 \label{eq:lambda3}
 \end{equation}
Unlike in the case of (\ref{eq:lambda2}), where the number of parameters is variable,
in (\ref{eq:lambda3}) the number of parameters $\{(\beta_{kj},\gamma_{kj})\}_{1\leq j\leq R_{k+}; 1\leq k\leq m}$ is fixed.
%
Below we show how (\ref{eq:lambda3}) can be looked upon as analogous to (\ref{eq:lambda2}) under appropriate modelling assumptions
involving Dirichlet process. 

\subsection{Modelling response surfaces using Dirichlet process}
\label{subsec:dp2}

\ctn{Bhatta06b} assume that for each $k$, the parameters $\theta_{k1},\ldots,\theta_{kR_{k+}}$ are samples from
some prior distribution $G_k(\cdot)$ on $\Re\times\Re^+$, where $G_k\sim\mathcal D(\alpha
G_0)$ is a Dirichlet process defined by $\alpha$, a positive scalar, and prior expectation $G_0(\cdot)$, a specified
bivariate distribution function over $\Re\times\Re^+$. 
In other words,
\[[\theta_{k1},\ldots,\theta_{kR_{k+}}\mid G_k]\sim iid\hspace{2.0mm}G_k\hspace{3.0mm}\mbox{for}\hspace{2.0mm}k=1,\ldots,m\]

\[\mbox{and for each}\hspace{3mm}k,G_k\sim\mathcal D(\alpha G_0);\hspace{3.0mm}\mbox{$G_k$ are
assumed to be independent.}\]
A crucial feature of the above modelling style concerns the discreteness
of the prior distribution $G_k$, given the assumption of Dirichlet process; that is, under these assumptions,
the parameters $\theta_{kj}$ are coincident with positive probability.
This is the property that \ctn{Bhatta06b} exploits to show that (\ref{eq:lambda3}) boils down to (\ref{eq:lambda2})
under the above modelling assumptions. The main points regarding this are sketched below.

Marginalisation over $G_k$ yields
\begin{equation}
[\theta_{kj}\mid\theta_{k1},\ldots,\theta_{k,j-1},\theta_{k,j+1},\ldots,\theta_{kR_{k+}}]\sim
\alpha a_{R_k
-1}G_0(\theta_{kj})+a_{R_{k+} -1}\sum_{l=1,l\neq
j}^{R_{k+}}\delta_{\theta_{kl}}(\theta_{kj})
\label{eq:prior}
\end{equation}
In the above,  $\delta_{\theta_{kl}}(\cdot)$ denotes a unit point mass at
$\theta_{kl}$ and $a_j=1/(\alpha +
j)$ for positive integers $j$.

The above expression shows that the
$\theta_{kj}$ follow a general Polya urn scheme, that is,
the joint distribution of $\{\theta_{k1},\ldots,\theta_{kR_{k+}}\}$ is given by the following:
$\theta_{k1}\sim G_0$, and, for $j=2,\ldots,R_{k+}$,
$\left[\theta_{kj}\mid\theta_{k1},\ldots,\theta_{k,j-1}\right]\sim\alpha a_{j-1}G_0(\theta_{kj})+a_{j-1}
\sum_{l=1}^{j-1}\delta_{\theta_{kl}}(\theta_{kj})$. 
Thus, given a sample $\{\theta_{k1},\ldots,\theta_{k,j-1}\}$, $\theta_{kj}$ is
drawn from $G_0$ with probability $\alpha a_{j-1}$ and is otherwise drawn uniformly
from among the sample $\{\theta_{k1},\ldots,\theta_{k,j-1}\}$. In the former case, $\theta_{kj}$ 
is a new, distinct realisation and in the latter case, it coincides with one of the realisations already
obtained. Thus, there is a positive probability of coincident values.
For more on the
relationship between a generalized Polya urn scheme and the
Dirichlet process prior, see \ctn{Blackwell73} and
\ctn{Ferguson73}.

Now, supposing that a sample from the joint
distribution of $\theta_{k1},\ldots,\theta_{kR_{k+}}$ yields $r_k$ distinct realisations
given by $\theta^*_{k1},\ldots,\theta^*_{kr_k}$, and if $R_{kj}$ denotes the number of times
$\theta^*_{kj}$ appears in the sample, then $R_{k1}+\ldots +R_{kr_k}=R_{k+}$. Hence, (\ref{eq:lambda3}) reduces
to (\ref{eq:lambda2}).

We remark that the prior for $r_k$ is implicitly
induced with this modelling style; for more details, see \ctn{Antoniak74}, \ctn{Escobar95}.

\subsection{Choice of $G_0$}
\label{subsec:base_measure}

To complete the Bayesian model description, it is necessary to specify the prior mean $G_0(\cdot)$ of
$G(\cdot)$. \ctn{Bhatta06b} assume that under $G_0(\cdot), {\gamma_{kj}}\sim
IG(11,30)$, an inverse-gamma prior with mean 3 and variance 1, and
$\left[\beta_{kj}\mid\gamma_{kj}\right]\sim
N(11.19,\frac{2.45}{9}{\gamma_{kj}}^2)$. Note that, since $E^2(\gamma_{kj})=9$, very roughly,
$Var(\beta_{kj})\approx 2.45$, which roughly corresponds to the prior of \ctn{Vasko00}.
We need to specify a value of $\alpha$. In order to do this reasonably, we adopt the following
elicitation arguments.
Note that the value of $\alpha$ is the one that approximately
(in a subjective sense) optimises the trade off between unimodal and multimodal components,
keeping in mind that {\it a priori} we expect the response surface to be multimodal. In other words,
it is necessary to reflect this ``optimism" about multimodality into the prior for the number of components.
With $\alpha=10$ and the maximum number of components, denoted by $R_{k+} = 10$ for each $k$, the 
probability of obtaining a distinct component is $\alpha/(\alpha+R_{k+}-1)=0.53$, which is slightly
more than the probability of obtaining a non-distinct component, the probability of the latter
being 0.47. Thus, {\it a priori},
our optimism about single component is slightly less than that about multiple components.
This seems reasonable, given prior palaeoclimatological knowledge. 
With this choice  \ctn{Bhatta06b} found that {\it a posteriori} the number of components of each species
was less than 10.


\subsection{Results of model assessment using IRD}

In the posterior analysis, the number of components
for each species was found to be greater than one with high probability, confirming that indeed the unimodal model
for relating species to environment is inappropriate. From the
cross-validation exercise with IRMCMC it was found that for this flexible multimodal modeling approach of \ctn{Bhatta06b}
82\% of the observed values fell within 95\% highest posterior density regions. This is a significant improvement over the unimodal modeling approach of \ctn{Vasko00}
where more than 40\% observed values were excluded from the 95\% highest posterior density regions. 
However, the goodness of fit test as described in this work was not satisfied, showing that there is scope to further improve the model. We reserve as future research the task
of further improving
the model until it satisfies the goodness of fit test.
Next we provide brief details of another much more complicated palaeoclimate problem.

\section{Brief discussion of goodness of fit test of a more difficult palaeoclimate problem}
\label{sec:pollen}

\ctn{Bhattacharya04b} provide inverse cross-validation analysis of the much more complicated palaeoclimate model
of \ctn{Haslett06}. In that case, pollen data was used, rather than chironomid data
and each cross-validation posterior involved 2 climate variables, 14 species of pollen 
and about 10,000 parameters. In all, there were 7815 cross-validation densities since there are as many cases.
Brute force MCMC implementation to explore all the 7815 cross-validation densities is expected to take about 5 years,
and this using highly sophisticated parallel computing architecture.
However, using IRMCMC, the entire exercise was completed in less than 8 hours.
For details on the implementation, see \ctn{Bhattacharya04b}.
Considering each climate variable singly, it was observed that the 95\% HPD credible regions of both the climate variables 
were rather large. Hence, although
in about 92\% and 97\% cases the true values were included within the respective HPD credible intervals of
the two climate variables, our goodness of fit test applied individually to the two climate variables
indicated that the model overfitted the data. One of the reasons that the model overfitted the data can be attributed 
to the presence of such a large number of unknown parameters
in the model and the use of vague priors. Moreover, a very large number of the cross-validation densities turned
out to be highly multimodal. \ctn{Bhatta06a} demonstrate by simulation study that lack of homogeneity
between different species is the reason for the multimodalities. In other words, since different species 
have different preferences for climate, the densities of climate variables were forced to be multimodal.
Clearly, such multimodalities, which result due to clashing information, increase the uncertainty about
climate variables.
For complete details on the assessment of fit of this pollen based palaeoclimate model, see \ctn{Bhattacharya04c}.


\newpage


\begin{table}
\caption{Forward problem: assessment of Poisson model fit when the true model is Geometric.}
\label{table:fwd_pg}
\begin{center}
\begin{tabular}{|c||c||c|}\hline
Parameter($\theta$) & Poisson agreement (\%) & Geometric agreement (\%)\\
\hline
0.1 & 97.0 & 99.7\\
1.0 & 59.7 & 99.0\\
3.0 & 13.2 & 98.8\\
5.0 & 3.3 & 98.5\\
7.0 & 0.8 & 99.1\\
15.0 & 0.0 & 97.5\\
\hline
\end{tabular}
\end{center}
\end{table}


\begin{table}
\caption{Inverse problem: assessment of Poisson model fit when the true model is Geometric.}
\label{table:inv_pg}
\begin{center}
\begin{tabular}{|c||c||c|}\hline
Parameter($\theta$) & Poisson agreement (\%) & Geometric agreement (\%)\\
\hline
0.1 & 97.3 & 97.3\\
1.0 & 89.3 & 97.1\\
3.0 & 63.5 & 97.5\\
5.0 & 34.7 & 97.7\\
7.0 & 18.4 & 97.8\\
15.0 & 1.4 & 97.6\\
\hline
\end{tabular}
\end{center}
\end{table}

%
%
\begin{table}
\caption{Assessment of inverse model fit.}
\label{table:inverse_model_fit}
\begin{center}
\begin{tabular}{|c||c|}\hline
Exponential mean($\lambda$) & Agreement percentage \\
\hline
0.5 & 56.0\\
1.00 & 74.4\\
3.00 & 90.4\\
10.00 & 95.4\\
\hline
\end{tabular}
\end{center}
\end{table}

\newpage


\bibliography{irmcmc}

\end{document}